\documentclass[aps,prd,nofootinbib,amsmath,amssymb,superscriptaddress,twocolumn,10pt]{revtex4}

\pdfoutput=1

\usepackage{graphicx}
\usepackage{dcolumn}
\usepackage{bm}
\usepackage{amssymb}
\usepackage{latexsym}
\usepackage{booktabs}
\usepackage{amsmath}
\usepackage{multirow}
\usepackage[colorlinks=true, linkcolor=red, citecolor=blue]{hyperref}

\newcommand{\be}{\begin{equation}}
\newcommand{\ee}{\end{equation}}
\newcommand{\bq}{\begin{eqnarray}}
\newcommand{\eq}{\end{eqnarray}}

\usepackage[usenames,dvipsnames]{xcolor}

\begin{document}

\title{A novel method of measuring cosmological distances using broad-line regions of quasars}

\author{Hong Li}
\affiliation{Key Laboratory of Particle Astrophysics, Institute of High Energy Physics, Chinese Academy of Sciences, Beijing 100049, China}
\author{Xin Zhang\footnote{Corresponding author}}
\email{zhangxin@mail.neu.edu.cn}
\affiliation{Department of Physics, College of Sciences, Northeastern
University, Shenyang 110819, China} 
\affiliation{Ministry of Education's Key Laboratory of Data Analytics and Optimization
for Smart Industry, Northeastern University, Shenyang 110819, China}
\affiliation{Center for High Energy Physics, Peking University, Beijing 100080, China}
\affiliation{Center for Gravitation and Cosmology, Yangzhou University, Yangzhou 225009, China}


\maketitle

Our ancestors, such as Cro-Magnon, always looked up at the stars and wished to understand the mysteries of the universe, but their view of the universe was probably quite simple: the stars are inlaid on the inner surface of the sky, and they are very small. However, now we know that stars and galaxies are actually extremely huge, and they look small because they are too far away from us. This indicates that the measurement of cosmological distance is always very important in the study of the universe. Due to the expansion of the cosmos, the stars we see are actually what they used to be. The wavelength of light emitted from a star becomes longer due to the expansion of the universe, which is called ``redshift of light''. The larger the redshift, the older is the star. If we can measure both the distances and redshifts of stars (luminous objects), we can then establish a relationship between distance and redshift, from which the information of cosmic expansion history can be obtained. Actually, the measurement of the expansion history of the universe can answer the fundamental questions about the cosmos, such as its age, geometry, composition, and so forth.

To measure the distance at a cosmological scale, two typical means are often used, called ``standard candle'' and ``standard ruler''. A type of supernovae, named type Ia supernovae (SNe Ia), are viewed as the standard candles in the universe, and {by observing them} the acceleration of the cosmic expansion was discovered. This great discovery led to the important notion that the universe is dominated by a mysterious component, dark energy, with repulsive gravity. Another way of measuring distance is provided by the standard ruler, the baryon acoustic oscillations (BAO), {which is a kind of relic imprints of primordial sound waves in the cosmic microwave background (CMB) and the distribution of matter (e.g., galaxies and neutral hydrogen).} 
The standard candle and the standard ruler measure the luminosity distance and the angular diameter distance, respectively. But, actually, the supernova observation does not measure the absolute luminosity distances, but the relative luminosity distances between supernovae at different redshifts (because the true luminosity of SN Ia is not exactly known). To measure the absolute luminosity distances {we need to use} the distance ladder provided by the observation of nearby universe to calibrate the low-redshift distances. The absolute distance measurements for the nearby universe are usually provided by the Cepheid observations, which can estimate the absolute luminosities of Cepheids through measuring their periods and then infer their absolute distances. The absolute luminosity distances of SNe Ia hosted along with Cepheids in the same galaxies can be calibrated by the measurements for the Cepheids. {In the actual construction of a distance ladder, the independent geometric calibrations of Cepheids are also important. For example, the recent improved geometric distance estimates to the Large Magellanic Cloud using detached eclipsing binaries and to the galaxy NGC 4258 using water masers have played a crucial role in the mission of building a distance ladder.} 
Of course, this distance-ladder method is affected by the extinction and reddening correction in the luminosity measurements of Cepheids and SNe Ia, and SNe Ia observations are also influenced by the standardization process of the Phillips relation.

The extremely precise measurements for the CMB anisotropies by the successive three generations of satellites, COBE, WMAP, and Planck, have pushed the studies for the cosmos into an era of ``precision cosmology''. The latest Planck observation \cite{Aghanim:2018eyx} has provided precise measurements for a few cosmological parameters, such as the densities of the components of the universe, the current expansion rate of the universe (namely, the Hubble constant), and the parameters related to the primordial perturbations generated from the infant universe. In particular, from a cosmological fit to the Planck CMB observation by assuming a cosmological standard model at present, the Hubble constant can be measured to the precision of about a few parts per thousand. The precision is fairly high, but the problem is that the method is model-dependent. 

It should be stressed that the Hubble constant is the first cosmological parameter proposed by cosmologists (actually, it was proposed by Edwin Hubble himself in the 1920s), and actually it has been measured for about one century. 

A traditional way of measuring the Hubble constant is to use the distance ladder for measuring the nearby-universe distances order by order. A drawback of this method is that for every rung of the ladder there may exist some poorly understood systematic errors. Currently, the precision of measuring the Hubble constant has reached about two percent, which is an amazing advancement for this method \cite{Riess:2019cxk}. 

Nevertheless, there appeared a severe situation that the results derived from the two methods, the CMB observation that is a measurement for the early universe and the distance-ladder observation that is a measurement for the late (nearby) universe, are in great discrepancy (namely, tension). The tension between them has reached the statistical confidence level of astonishing 4--6 standard deviations. The Hubble constant is very important for cosmology because it is involved in all the calculations concerning cosmological distance and time in the theoretical cosmology, and thus precisely measuring the Hubble constant is of extreme significance. 
{More importantly, the local measurement of the Hubble constant provides a crucial end-to-end test for the current ``standard model of cosmology'', called $\Lambda$CDM (lambda cold dark matter). }
Therefore, the ``Hubble tension'' has led to a severe crisis for cosmology \cite{Verde:2019ivm,Riess:2020sih}. 
{Since the current estimate of $H_0$ from CMB depends on the $\Lambda$CDM model, if both the local and CMB measurements are confirmed, this will imply that the standard $\Lambda$CDM model is not correct, and that some ``new physics'' has to be invoked to fix the tension on $H_0$.}


To solve the crisis, the first step is to find a method of measuring the nearby-universe distances that should be independent of a distance ladder and of a cosmological model. Such an independent measurement could provide a third-party arbitration for the current ``Hubble tension''. Thanks are given to active galactic nuclei (AGNs) and quasars for such a unique opportunity. Their spectra are characterized by broad emission lines, which are from ionized clouds in the so-called broad-line regions (BLR). These celestial objects are just siting a position  allowing us to conveniently measure linear (physical) and angular sizes of the BLRs of AGNs and quasars through reverberation mapping (RM) technique and spectroastrometry. Fortunately, a geometric way of precisely measuring cosmological distances can be achieved by a joint analysis. The first efforts have been made by a Chinese team  through  GRAVITY/VLTI and RM campaign.

AGNs actually refer to supermassive black holes (SMBHs) in the centers of galaxies, which are one of the brightest objects in the universe. Understanding the physics of BLR of SMBH is helpful in measuring the basic parameters of black holes, such as black hole mass, linear angular scale, etc., which plays an important role in the study of SMBHs. In the past few decades, RM has become the main method to study the kinematics and geometry of BLR and to measure the mass of SMBH as AGN, and great progress has been made in this aspect. It identifies the BLRs in time domain, instead of spatially, by monitoring spectroscopically the response of broad emission lines to the variations of continuum fluxes.

Since 2012, through joint observations by the Bok 2.3-meter telescope in Steward Observatory, University of Arizona, and the Lijiang 2.4-meter telescope in Yunnan Observatory, Chinese Academy of Sciences, the RM continuous observations for more than 50 AGNs are performed and their time delays between the variations of continuum and emission lines are also successfully measured, providing better understandings of the super-Eddington accretion onto black holes, the BLR physics, and their potential as a new probe of cosmological distances. By means of spectroastrometry (SA), the terminal instrument, GRAVITY, equipped at the Very Large Telescope Interferometer (VLTI), can achieve a spatial resolution of up to 10 microseconds in the near infrared band. It can successfully reveal the structure, kinematics, and angular size of the BLR of the AGN 3C 273 with an unprecedentedly high spatial resolution, and it can provide the measurements of the angular structure of BLR in a direction perpendicular to the line of sight. {RM observations provide the physical linear sizes of BLRs,} and thus a joint analysis of SA and RM observations of AGNs can directly measure absolute angular diameter distances ($D_A$). This measurement method is a purely geometrical measurement without any physical assumptions, thus it does not rely on any existing distance ladders, nor on the corrections to light extinction, reddening and standardization necessary for traditional tools.  As it measures the angular diameter distance to AGN directly, this method can avoid some of uncertainties that inhabited in the traditional distance measurement ways reasonably and systematically.

With this idea, recently, writing in {\it Nature Astronomy}, Wang et al. \cite{Wang:2019gaq} reported a direct measurement on the absolute $D_A$ of AGN 3C 273, the first quasar discovered at the redshift of 0.158, and gave its cosmological indication for $H_0$. The result is very meaningful for cosmology, as it is totally different from the local distance-ladder measurement in that it does not rely on the order-by-order measurements of local distances, avoiding the cumulated systematics caused by the local distance ladder. Of course, in the method, it is no need to assume a cosmological model. Therefore, it is an independent distance measurement.

Measuring $D_A$ with SMBH, such a research has proposed a new way for measuring the cosmic distance and provided a new, independent measurement for the Hubble constant. Due to the limitation of only one event being observed, the measurement value of $H_0$ given from the observation of 3C 273 is  $71.5_{-10.6}^{+11.9}$  km s$^{-1}$ Mpc$^{-1}$, and one can see that the current uncertainty is about 15\%. However, it is worth noting that the quasars like 3C 273 are not unique in the universe, and according to the current observational ability of GRAVITY, {its observational sample is going to increase.} Therefore, the cooperative observation of GRAVITY-RM is expected to further improve the measurement accuracy of $H_0$ in the forthcoming years, providing independent and accurate measurement of $H_0$ to help solve the ``$H_0$ crisis''. 
It should be noted that there is a project in progress to make measurements for 50 AGNs, for which the measurements in the redshift range of $0.5<z<2$ will be a big challenge, because the relevant RM observation needs longer time due to the time dilation effect of $(1+z)$ and the mid-scale telescope with a 4--5 m aperture is also needed.

In fact, recently, some other promising new cosmological probes have been discussed. For example, in the circumstance of the first ever image of the supermassive black hole M87$^{\ast}$ being recently captured by the Event Horizon Telescope, the possibility of using the shadows of supermassive black holes as a new standard ruler in the study of cosmology has been recently discussed and investigated \cite{Tsupko:2019pzg,Qi:2019zdk}. In addition, the multi-messenger observations for the event of binary neutron star merger (GW170817) have shown that the future gravitational wave (GW) observations would play a crucial role in the study of cosmology. In the multi-messenger observations involving the observation of GWs, the sources of GWs in this type of observations are also called cosmological ``standard sirens'', by which a true distance--redshift relation can be established. Since the waveform of GW encodes the information of absolute luminosity distance, the GW observations can provide an absolute distance measurement without using the local distance ladder. The analysis for GW170817 as a standard siren gave an independent measurement of the Hubble constant, with the precision of about 15\% \cite{Abbott:2017xzu}, and a further forecasting analysis has shown that about 50 similar standard sirens can achieve about 2\% measurement for the Hubble constant, similar to the precision of the current distance-ladder result, which will be realized in the near future \cite{Chen:2017rfc}. Therefore, the future GW standard siren observations from the third-generation ground-based detectors, e.g., Einstein Telescope and Cosmic Explorer, and the space-based detectors, e.g., LISA, Taiji, and TianQin, will definitely become a new powerful cosmological probe (see Ref.~\cite{Zhang:2019ylr} for a brief review). It is expected that these new distance measurement methods would play important roles in helping answer fundamental questions in cosmology in the future.



\begin{acknowledgments}

This work was supported by the National Natural Science Foundation of China (Grant Nos. 11975072,  11690021, 11875102, 11835009 and 11653002), the Liaoning Revitalization Talents Program (Grant No. XLYC1905011), the Top-Notch Young Talents Program of China, the Youth Innovation Promotion Association Project of the Chinese Academy of Sciences, and the Strategic Priority Research Program of the CAS (Grant No. XDB23020000).


\end{acknowledgments}



\end{document}